# Pathological MRI Segmentation by Synthetic Pathological Data Generation in Fetuses and Neonates

*Misha P.T Kaandorp[1,2], Damola Agbelese[1], Hosna Asma-ull[3], Hyun-Gi Kim[3], Kelly Payette[1,4], Patrice Grehten[5], Gennari Antonio Giulio[1,6], Levente István Lánczi[7], Andras Jakab[1,2]*

[1] Center for MR Research, University Children's Hospital Zurich, Zurich, Switzerland
[2] University of Zurich, Zurich, Switzerland
[3] Department of Radiology, Eunpyeong St. Mary's Hospital, College of Medicine, The Catholic University of Korea, Seoul, Republic of Korea
[4] Department of Early Life Imaging, School of Biomedical Engineering and Imaging Sciences, King's College London, London, UK
[5] Department of Diagnostic Imaging, University Children's Hospital, Zurich, Switzerland
[6] Department of Neuropediatrics, University Children's Hospital Zurich, Zurich, Switzerland
[7] Clinical Center Medical Imaging Clinic, University of Debrecen, Debrecen, Hungary

[*] **Corresponding Author**

**Name:** Misha Pieter Thijs Kaandorp

**Department:** Center for MR Research

**Institute:** University Children's Hospital Zurich

**Address:** Lenggstrasse 30, 8008 Zürich

**Email:** Misha.Kaandorp@kispi.uzh.ch

**Abstract**

Developing new methods for the automated analysis of clinical fetal and neonatal MRI data is limited by the scarcity of annotated pathological datasets and privacy concerns that often restrict data sharing, hindering the effectiveness of deep learning models. We address this in two ways. First, we introduce Fetal&Neonatal-DDPM, a novel diffusion model framework designed to generate high-quality synthetic pathological fetal and neonatal MRIs from semantic label images. Second, we enhance training data by modifying healthy label images through morphological alterations to simulate conditions such as ventriculomegaly, cerebellar and pontocerebellar hypoplasia, and microcephaly. By leveraging Fetal&Neonatal-DDPM, we synthesize realistic pathological MRIs from these modified pathological label images. Radiologists rated the synthetic MRIs as significantly ($p < 0.05$) superior in quality and diagnostic value compared to real MRIs, demonstrating features such as blood vessels and choroid plexus, and improved alignment with label annotations. Synthetic pathological data enhanced state-of-the-art nnUNet segmentation performance, particularly for severe ventriculomegaly cases, with the greatest improvements achieved in ventricle segmentation (Dice scores: 0.9253 vs. 0.7317). This study underscores the potential of generative AI as transformative tool for data augmentation, offering improved segmentation performance in pathological cases. This development represents a significant step towards improving analysis and segmentation accuracy in prenatal imaging, and also offers new ways for data anonymization through the generation of pathologic image data.

**Keywords**: Generative AI, Synthetic MRI, Diffusion models, Fetal and neonatal MRI, Segmentation performance

## 1. Introduction

Fetal and neonatal magnetic resonance imaging (MRI) are increasingly valuable diagnostic tools for studying neurodevelopment during the perinatal period and play an important role in the clinical counseling of congenital and acquired diseases (Benkarim et al., 2017; Jakab et al., 2021; Lautarescu et al., 2024). Prenatal development encompasses dynamic morphological changes in the brain, which continue to evolve throughout infancy. Pathologies that form during this time can significantly disrupt brain structure and development with long term implications. Fetal and neonatal MRI allows for both qualitative and quantitative analyses to identify healthy and pathological patterns (Pfeifer et al., 2019; Sled and Nossin-Manor, 2013). Despite its potential, developing automated Artificial Intelligence (AI) pipelines for fetal and neonatal MRI analysis that improve standardization, efficiency, and reliability while minimizing human errors remains challenging, especially in pathological conditions that substantially affect brain structures. Unlike the well-established frameworks for adult MRI, progress in this domain faces challenges due to the unique and dynamic anatomy of the developing brain, data heterogeneity, limited availability of expert annotated datasets, and the low quality of acquired data (Dubois et al., 2021; Jarvis and Griffiths, 2019; Sanchez et al., 2024).

Prior to the advent of deep learning (DL), several atlas-based methods were developed for brain tissue segmentation (Gholipour et al., 2012; Habas et al., 2010), but their applicability was restricted to normally developing fetuses. The Fetal Tissue Annotation (FeTA) MICCAI challenges (Payette et al., 2024, 2023, 2021) demonstrated that DL models can achieve robust performance in multi-label segmentation of T2-weighted (T2w) 3D super-resolution reconstructed (SR) fetal MRI. However, while these DL models perform well on healthy subjects, their performance drops when tested on diverse pathological cases, such as those with ventriculomegaly. Recent advancements, including DL-based segmentation algorithms such as BOUNTI (Uus et al., 2023) and FetalSynthSeg (Zalevskyi et al., 2024), have shown promising results in segmenting healthy fetal brain MRI but also underscore persistent challenges in handling pathological data. The available data for training image segmentation networks for the development of fetal and neonatal algorithms is limited. First, MRI annotation by manual delineation of structures requires extensive resources. Second, data availability for all developmental stages and different pathologies is limited. Third, data privacy regulations may restrict open data sharing in cases where the presence of rare pathologies might facilitate subject re-identification.

Recent advancements in generative AI have significantly enhanced the ability to generate realistic images, with Generative Adversarial Networks (GANs) (Goodfellow et al., 2014; Kazeminia et al., 2020; Skandarani et al., 2023) and diffusion models to (Croitoru et al., 2023; Kazerouni et al., 2023) being central these advancements. GANs are a class of deep learning models that generate data by training two neural networks—the generator and the discriminator—in an adversarial training process. GANs have shown success in creating synthetic images; however, they are prone to instability, leading to issues like mode collapse, vanishing gradients, and convergence problems. Moreover, many GAN techniques are primarily designed for two-dimensional (2D) images, which falls short of meeting the three-dimensional (3D) data needs in medical imaging (Kazeminia et al., 2020; Skandarani et al., 2023). A typical method involves generating 2D slices and stacking them to form 3D images, which can introduce spatial inconsistencies and overlook essential 3D contextual information (Jung et al., 2021). For instance, a previously reported SPADE

GAN (Spatially Adaptive Denoising GAN) implementation for fetal MRI (Fernandez Garcia et al., 2022) showed promising results but faced limitations in generating fully realistic images and suffered from interslice artifacts due to its 2D nature. In contrast, diffusion models, such as Denoising Diffusion Probabilistic Models (DDPMs) (Ho et al., 2020) offer several advantages over GANs. Unlike GANs, diffusion models progressively refine noise into a coherent image, producing higher-quality and more stable results (Croitoru et al., 2023; Dhariwal and Nichol, 2021; Kazerouni et al., 2023). One such model, Med-DDPM (Dorjsembe et al., 2024) was recently proposed for 3D semantic brain MRI synthesis. Serving as a substitute for real data, the use of synthetic data as training data for deep learning models have been increasingly utilized in medical image analysis applications, such as model fitting (Kaandorp et al., 2025, 2023) and image segmentation (Billot et al., 2023; Gao et al., 2022; Zalevskyi et al., 2024).

In this work, we propose a novel generative AI framework for generating realistic synthetic fetal and neonatal pathological MRIs from morphologically modified semantic label images. We aim to enrich data by the generation of realistic synthetic fetal pathological MRIs derived from healthy MRIs while ensuring that the generated images remain clinically realistic. To validate our approach, experienced radiologists assess the quality of the images focusing on the clinical diagnostic value of the synthetic images. In the context of more recent initiatives in this research domain, we evaluate how the proposed method impacts segmentation accuracy using published benchmark datasets and in-house clinical datasets of fetal pathologies. Moreover, we validate the generalizability of our approach in out-of-distribution data using a geographically diverse dataset.

## 2. Methods

### 2.1. In vivo data

This study utilized retrospective T2w cerebral MRI data from prior research. A total of six datasets were included, comprising both retrospectively sampled clinical data and publicly available datasets, which are described in detail below. Table 1 summarizes the imaging and preprocessing characteristics of each dataset. Table 2 (*left*) provides an overview of the number of cases (*neurotypical, pathological*) and age ranges for each dataset.

The corresponding local ethics committees independently approved the studies under which data were collected, and all participants gave written informed consent. In case of the data collected in Zürich, the Cantonal Ethical Committee of Zürich approved the study (Decision numbers: 2016-01019 and 2022-01157).

#### 2.1.1. FeTA2021 dataset

The FeTA2021 dataset, originally created for the Fetal Tissue Annotation and Segmentation (FeTA) 2021 MICCAI challenge (Payette et al., 2023), comprises of 120 fetal MRI brain scans acquired at the University Children's Hospital Zurich, Zurich, Switzerland. Of these, 80 scans were designated for the training set (FeTA2021$_{train}$), available on Synapse (Payette and Jakab, 2021), while the remaining 40 scans were part of the test set (FeTA2021$_{test}$), which is not publicly accessible. This followed the same split used in the FeTA2021 challenge. Both the training and test sets include neurotypical and pathological cases, with the pathologies including conditions such as Chiari-II malformation in spina bifida with ventriculomegaly, however, a more detailed break-down of pathologies is not possible due to dataset anonymization. Subjects in both FeTA2021$_{train}$ and FeTA2021$_{test}$ ranged from 21 to 35 weeks of gestational age (GA). T2w single-shot Fast Spin Echo (ssFSE) images were acquired in three planes with a resolution of 0.5×0.5×3-5 mm, using 1.5T or 3T GE MRI scanners (Signa Discovery MR450 and MR750). Imaging parameters included TR: 2000-3500 ms, TE: 120 ms, flip angle: 90°, field of view: 200–240 mm, and a sampling percentage of 55%. Further acquisition details are found in (Payette et al., 2021) and Payette et al. (2023).

Super-resolution (SR) reconstructions were performed using mialSR for 60 cases and Simple IRTK for another 60 cases, producing isotropic volumes with a resolution of 0.5 mm$^3$. Reconstructed images were zero-padded to dimensions of 256×256×256 voxels. Maternal tissue was excluded from the SR reconstructions, only the fetal brain was reconstructed. Each case included a 3D reconstruction and a manually segmented 7-label image, which included: white matter, gray matter, external cerebrospinal fluid (CSF), ventricles, brainstem, cerebellum, and deep gray matter. More description on the dataset structure and annotations is given in the original publications (Payette et al., 2023, 2021; Payette and Jakab, 2021).

### 2.1.2. Zurich-spinabifida

The Zurich-spinabifida dataset, initially described in (Payette et al., 2019) and later expanded consists of clinical fetal MRI scans from 90 subjects with spina bifida who underwent prenatal surgical repair. Age range was 25 to 30 weeks GA. MRI acquisition parameters matched those used in the FeTA2021 dataset. Original scans were processed using a semi-automated MeVisLab module for reorientation and masking. SR reconstruction (Tourbier et al., 2015) was applied to each set (3 to 14) of images taken in various orientations (axial, sagittal, coronal) for each subject, resulting in 3D volumes that represented fetal brain. Maternal tissue was excluded from the images, and each case included a 3D reconstruction and a semi-automatically generated segmented 7-label image.

We compiled a separate clinical dataset comprising the 26 most severe ventriculomegaly cases within the spina bifida group (Zurich-spinabifida$_{severe}$), selected based on qualitative assessments, with an age range of 25 to 30 weeks GA.

### 2.1.3. Zurich-controls

The Zurich-controls dataset includes 44 neurotypical fetal MRI scans acquired at the University Children's Hospital Zurich in Zurich, Switzerland. Subjects ranged in age from 20 to 34 weeks GA. Imaging parameters were consistent with those used in FeTA2021. Original scans were processed using a semi-automated NeSVoR module (Xu et al., 2023) for slice-to-volume reconstruction (SVR). Maternal tissue was excluded, and each case included a 3D reconstruction and a semi-automatically generated 7-label segmented image.

### 2.1.4. CU-neonates

The Catholic University of Korea-neonates (CU-neonates) dataset consists of 94 preterm-born neonatal MRI scans acquired at The Catholic University of Korea, Eunpyeong St. Mary's Hospital, South Korea, during their Neonatal Intensive Care Unit (NICU) admission. Age range was 29 to 42 weeks gestational corrected age (GCA). The neonates were imaged using 3T MRI scanners (Magnetom Vida, Siemens Healthineers). For acquiring the T2w images, a 3D T2-weighted Sampling Perfection with Application-Optimized Contrasts using different flip angle Evolution (SPACE) was employed using the following scanning protocol: repetition time msec/echo time msec, 4000/562; and 0.8 mm isotropic voxel; variable flip angle. No further preprocessing was performed. Additional information on the CU-neonates dataset can be found in Park et al. (2024).

### 2.1.5. The developing Human Connectome Project

The developing Human Connectome Project (dHCP) is a research initiative that provides MRI datasets of fetal and neonatal brains to study early brain development. It includes MRI data from fetuses (dHCP fetal) and neonates (dHCP neonates), aimed at studying early brain development. Imaging was conducted at the Evelina Newborn Imaging Centre, St Thomas' Hospital, London, UK.

#### 2.1.5.1. dHCP neonates

This dataset (2nd release of dHCP neonates) includes 558 T2w scans from 505 neonates ranging in age from 23 to 44 weeks GCA. Specifically, the dataset contains 378 scans of term-born neonates and 180 scans of preterm-born neonates, of which 82 scans are from very preterm-born neonates (birth age < 32 gestational weeks). Imaging was performed on a 3T Philips Achieva scanner equipped with a 32-channel neonatal head coil. Scans were acquired during natural unsedated sleep following feeding. Imaging parameters included in-plane resolution of 0.8×0.8 $mm^2$ with 1.6 mm slices overlapping by 0.8 mm, TR: 12 s, TE: 156 ms, and SENSE factors of 2.11 (axial) and 2.60 (sagittal). Motion-corrected slices were reconstructed into 3D volumes using SVR. Anatomical segmentations included 9 tissue types and 87 regions using the dHCP neonatal pipeline. Further details on dHCP neonates structural imaging and preprocessing are described in Makropoulos et al. (2018) and Fitzgibbon et al. (2020).

#### 2.1.5.2. dHCP fetal

This dataset includes 297 T2w scans acquired using a 3T Philips Achieva system with a 32-channel cardiac coil, with subjects ranging from 21 to 38 weeks GA. Structural T2w imaging involved six uniquely oriented stacks centered on the fetal brain, acquired using a zoomed multiband single-shot TSE sequence with an MB tip-back preparation pulse to enhance SNR efficiency. SVR was used to produce isotropic 3D volumes, with automatic rejection of corrupted data (Jiang et al., 2007; Kuklisova-Murgasova et al., 2012). Imaging protocols also addressed field inhomogeneities and motion artifacts through B0 and RF shimming (Gaspar et al., 2019), local power scaling, and calibration with SENSE references, dual-TE B0, and DREAM (Nehrke and Börnert, 2012) B1 maps. The final datasets were reconstructed into isotropic 3D using SVR, incorporating automatic rejection for corrupted data. Anatomical segmentations were generated with 9 tissue types and 87 regions using BOUNTI (Uus et al., 2023). Further details on dHCP fetal structural imaging and preprocessing are described in (Price et al., 2019).

### 2.1.6. Further preprocessing

All datasets, except for the CU-neonates, included corresponding label images. For the CU-neonates, label images were generated using the dHCP neonate pipeline. The dHCP fetal dataset occasionally contained poor segmentations with random pixels within certain labels. To address this, connected components containing fewer than 20 pixels were reassigned to the label most frequently surrounding them.

### 2.2. Proposed framework

Our proposed framework comprised of two main steps : diffusion model training (Figure 1A) and label image modifications together with pathological fetal and neonatal MRI synthesis (Figure 1B).

#### 2.2.1. Diffusion model training

##### 2.2.1.1. Preprocessing

Further preprocessing to the in vivo data were applied for diffusion model training. We considered all datasets for training, except the CU-neonates. We hypothesized that excluding poor-quality data from the training data would enhance the model's ability to generate higher-quality synthetic images. Consequently, we excluded 650 images from the combined fetal dataset based on qualitative assessments of motion artifacts or low resolution, resulting in 727 final fetal images. Neonatal images were not excluded, as all met high quality. The age range of the training set was 21 to 44 weeks GCA. Table 2 (*right*) provides detailed information about the number of cases (*neurotypical, pathological*) and the specific age ranges for each dataset included in the diffusion model training.

After quality filtering, original segmented label images were mapped into the same four classes: Class 0 for the background, Class 1 for brain fluid (including external cerebrospinal fluid and lateral ventricles), Class 2 for the cortex (gray matter), and Class 3 for miscellaneous structures (such as the brainstem, cerebellum, deep gray matter, and white matter). The MRI-label pairs were cropped to remove excess background. We performed intensity scaling by normalizing intensity ranges to [-1, 1] as described in Dorjsembe et al. (2024a), and isotropically resized the images to 160×160×160 before being input into the network. Note that resizing to larger image dimensions (larger than $160^3$) resulted in computational constraints during diffusion model training (NVIDIA RTX A6000 48GB GPU).

##### 2.2.1.2. Fetal&Neonatal-DDPM

The diffusion model to generate synthetic pathological fetal and neonatal MRIs from modified label images builds upon the Med-DDPM framework described by Dorjsembe et al. (2024a). This framework uses DDPMs (Ho et al., 2020), which applies random noise to real images (forward diffusion process) and train a DL model to iteratively reverse this noise through a reverse denoising process. In Med-DDPM the denoising is guided by conditioning the denoising step with semantic label images through channel-wise concatenation with the input image.

We modified Med-DDPM to accommodate four input classes (instead of 3) by adding an extra channel in the concatenation step. We altered the original hyperparameters as described in Dorjsembe et al. (2024a) to improve performance. Specifically, we increased the step size to 1000 steps (instead of 250) and set the number of epochs to 500,000 (instead of 100,000). The reasons for this was that fetuses and neonates have higher anatomical variability, compared to the more consistent adult brain structures. These adjustments enhanced network stability and improved the realistic nature of the synthetic images. During the first 100,000 epochs, a learning rate of $10^{-5}$ was used, which was subsequently decreased to $10^{-6}$ for the remaining 400,000 epochs. Other hyperparameters, consistent with those outlined by Dorjsembe et al. (2024a), include the L1 loss and the Adam optimizer using an Exponential Moving Average (EMA) strategy, with a decay factor of 0.995, to ensure stable and efficient training. This customized model, referred to as Fetal&Neonatal-DDPM, was trained on all 727 high-quality fetal and neonatal images. In addition to its applicability for neonatal MRI synthesis, neonatal data was incorporated to address the limited sample size of fetal data. Including neonatal data was motivated by ensuring the continuity of brain morphological characteristics across development, and to enhance both performance and stability. Furthermore, this way our Fetal&Neonatal-DDPM is also applicable for neonatal MRI synthesis.

### 2.2.2. Label image modification

To simulate pathological conditions, label images of healthy subjects were altered through label morphological modifications. These were specifically designed for ventriculomegaly, cerebellum and pontocerebellar hypoplasia, and global cerebral atrophy combined with microcephaly and extended extracerebral fluid spaces (referred to as 'atrophy/microcephaly' further in the manuscript), as well as their combinations. Details of each morphological modification, including the pathology generator used, are described below.

#### 2.2.2.1. Pathology generator

A generator was designed to create an infinite variety of pathological label images based on original healthy label images. This generator assigned a random severity level to each pathology, which was then used to modify the original label images. Determining the specific pathology for each modified pathological label image from healthy ones was carried out in sequential steps. Initially, each pathology had an equal probability of being simulated. Once a pathology was assigned, subsequent conditions were introduced with a 50% probability. However, cerebellum hypoplasia and pontocerebellar hypoplasia were mutually exclusive, and whenever atrophy/microcephaly was present then ventriculomegaly always co-occurred (but not the other way around) due to their frequent co-occurrence in patients with microcephaly. Furthermore, for ventriculomegaly, there was a 50% chance of simulating either symmetrical or asymmetrical ventricular dilation.

#### 2.2.2.2. Ventriculomegaly synthesis

Ventriculomegaly (Alluhaybi et al., 2022) is a common condition seen in fetal MRI, characterized by the enlargement of the brain's ventricles, the fluid-filled cavities that produce and store cerebrospinal fluid. Ventriculomegaly on fetal MRI, as well as its association with further abnormalities (for example in spina bifida), has important diagnostic and therapeutic consequences. (Jakab et al., 2021b). In this work, we refer to dilated lateral ventricles as ventriculomegaly.

Ventriculomegaly was simulated by dilating the lateral ventricular labels. This was done both symmetrically and asymmetrically between the two cerebral hemispheres. As the FeTA2021 label images only contained 7 labels without separation between the hemispheres, we first separated the white matter and ventricles labels in the midline. This hemisphere separation was achieved using an in-house tool that is based on K-Means clustering that divides the voxel coordinates into two (L-R) clusters based on the spatial location. This tool requires the brain's antero-posterior anatomical axis to be aligned parallel to the Y-axis of the image, so misaligned images were rotated for processing and reverted afterward using manual rotation. Once the ventricles and white matter labels were split between the hemispheres, we determined the maximum number of dilations allowed for the ventricle labels to fill 65% of the white matter volume in the corresponding hemisphere, to avoid the ventricle to reach the cortex in an unrealistic way. A random number of dilations was then assigned to each ventricle, constrained by this maximum. The ventricle labels were subsequently dilated within their respective hemispheres. The dilated ventricle labels maintained a minimum 2-pixel distance from other structures to avoid unrealistic overlaps and were smoothed. Finally, the dilated ventricle labels from both hemispheres were combined to create a single synthetic ventriculomegaly label image.

### 2.2.2.3. Hypoplasia synthesis

Hypoplasia describes the underdevelopment or incomplete development of an organ or tissue. In the context of the brain, hypoplasia typically refers to the underdevelopment of structures like the cerebellum (Poretti et al., 2014) or both cerebellum and brainstem (pontocerebellar) (Rudnik-Schöneborn et al., 2014).

Pontocerebellar hypoplasia was simulated by shrinking the brainstem, 4th ventricle, and cerebellum labels, using the centroid of these combined labels. Initially, the brainstem, cerebellum, and 4th ventricle labels were randomly shrunk in the x-y plane, with a maximum shrinkage of approximately 20% of the original brainstem size. The cerebellum and 4th ventricle labels were also reduced within these planes to prevent disconnection from the brainstem. As the FeTA2021 label images only include 7 labels without separation between the 4th ventricle and the lateral ventricle, the 4th ventricle was identified by performing 4-pixel dilations in the x-y direction and selecting the ventricular labels within the dilated mask. After shrinking the brainstem, cerebellum, and 4th ventricle labels in the x-y plane, the cerebellum was dilated using the same scaling factor as the shrinkage to restore its 'rounded' shape. Following this, the cerebellum label was further shrunk in all three planes (x, y, z) with the same scaling, ensuring proportional shrinkage in the z-plane as well, and shrunk around the point where the cerebellum attaches to the brainstem. For simulating cerebellar hypoplasia, only the cerebellum label was shrunk in all three planes.

### 2.2.2.4. Microcephaly and atrophy synthesis

Microcephaly (Leibovitz and Lerman-Sagie, 2018) is a condition where an individual has a smaller-than-normal head size, typically due to abnormal brain development. Here in this work, we referred to global brain atrophy with increased extracerebral spaces (e.g. associated with Zika encephalitis) as microcephaly.

Microcephaly was simulated by reducing the size of the entire brain labels within the external cerebrospinal fluid spaces. Initially, all labels were uniformly shrunk toward the centroid of all labels, with a maximum shrinkage of 10%. The space surrounding the shrunken brain that is within the original label map was then filled with external cerebrospinal fluid. As a result, the overall brain size was reduced.

### 2.2.3. Pathological fetal and neonatal MRI synthesis

The modified pathological label images underwent preprocessing as described in Section 2.1. Specifically, the label images were cropped, rescaled to [-1, 1], and isotropically resized to dimensions of 160×160×160. We synthesized MRI scans using both original and simulated semantic pathological labels from healthy subjects across three datasets. For the FeTA2021 challenge dataset, label images from 33 healthy cases of the training set (FeTA2021$_{train}$) were modified and corresponding synthetic MRIs were generated using Fetal&Neonatal-DDPM. For the dHCP fetal dataset, label images for 265 cases with available accurate label images were modified and corresponding MRIs were synthesized. To demonstrate our approach in out-of-distribution data, MRIs were synthesized for all CU-neonates subjects using original labels.

### 2.3. Evaluation

This section summarizes two evaluations: qualitative assessment of synthetic MRI quality by radiologists and segmentation performance evaluation using synthetic data in training state-of-the-art nnUNet models.

#### 2.3.1. Qualitative evaluation

Four board certified radiologists (Table 3) blinded to the image sources evaluated the clinical diagnostic value of these images. To prevent bias, the radiologists were informed that the objective was to evaluate the quality of a novel image-processing method, without revealing that some cases involved synthetic data. The evaluation included 100 cases: 50 real MRIs and 50 synthetic MRIs from fetal datasets FeTA2021 and dHCP fetal datasets, in random order. The real MRIs included 25 healthy cases and 25 pathological cases. The synthetic MRIs included 16 healthy cases, 17 pathological cases generated from healthy MRIs, and 17 cases based on real pathological MRIs. None of these evaluation cases were part of the diffusion model's training set. For simplicity, we provided the radiologist with one representative slice from each plane (axial, coronal, sagittal) for each case. The experts assigned one of four quality scores (0-3) to the data:

- **(0) Unusable quality:** The image is *unusable*, with significant artifacts, noise, or distortions rendering it unsuitable for diagnostic or interpretative purposes.
- **(1) Poor quality**: The image is *marginally diagnostic* but suffers from noise and artifacts that obscure structures.
- **(2) Good quality**: The image is *useful for most diagnostic purposes*, with sufficient clarity and resolution to interpret and delineate key structures such as ventricles, gray matter and white matter, cerebellum and brainstem though lacking the detail necessary for complex or nuanced evaluations. There might be artifacts present but does not limit the delineation of identification of structures.
- **(3) Excellent quality**: The image is *fully diagnostic and optimal*, with exceptional clarity, resolution, and contrast, enabling detailed analysis of both subtle and complex features. None or very subtle artifacts are present, but not interfere in assessing the diagnostic value.

Example images and additional information on quality scoring provided to the experts are available in Supplementary Information. The evaluations were based solely on visual assessments. We determined the average between all raters for the real and synthetic dataset and performed a two-sided unequal variance student t-test to evaluate statistically significant ($p < 0.05$) difference between the real and synthetic dataset utilizing the mean of each case.

To demonstrate generalizability of our approach to images on further cohorts including geographic diversity, we also qualitatively assessed synthetic MRIs generated from the CU-neonates dataset.

### 2.3.2. Segmentation performance

To evaluate whether fetal MRI segmentation performance is improved by adding synthetic pathological MRIs, we trained state-of-the-art nnUNet (v2) (Isensee et al., 2021) models on different configurations of training data. These configurations included both original and synthetic pathological MRI-label pairs derived from healthy label images. These configurations were derived both from FeTA2021$_{train}$ and dHCP fetal datasets to explore the utilization of real and synthetic data, small and large synthetic sample sizes, use of different base datasets (FeTA2021$_{train}$ vs. dHCP fetal) for generating synthetic pathological data, as well as synthetic data of multiple pathologies. We also assess the generalizability across structures. For training, the synthetic images were reverted to their original image dimensions and intensities of the derived healthy subjects. Base full-resolution 3D nnU-Net configuration settings were used, and training was conducted for 100 epochs.

Table 4 shows all 12 configurations of training data. Our experiments included 8 different configurations of simulated pathological MRI-label pairs generated from 33 healthy subjects in the FeTA2021$_{train}$ dataset and 265 healthy subjects in the dHCP fetal dataset (this is the number of healthy or neurotypically developing subjects in the original datasets we sampled). Baseline performance was established using nnU-Net models trained on original healthy MRI-label pairs: R33-healthy for FeTA2021$_{train}$ and R265-dHCP for dHCP fetal. Performance was also assessed for 80 healthy and pathological MRI-label pairs from FeTA2021$_{train}$ (R80-FeTA) and their synthetic equivalent (S80-FeTA). Here, 'R' refers to real data, and 'S' refers to synthetic data.

We measured Dice score performance on FeTA2021$_{test}$ (see Section 2.2.1) and Zurich-spinabifida$_{severe}$ (see Section 2.2.1), with age range 25 to 30 weeks GA. Overall performance of each configuration was assessed by averaging the median scores across all labels. Additionally, we qualitatively evaluated the nnU-Net performance through visual inspection of the segmentations.

### 3. Results

#### 3.1. Qualitative evaluation

Figure 2 shows a subset of synthetic pathological fetal MRIs from modified healthy MRI label images of both FeTA2021 Challenge (Figure 2A) and dHCP fetal (Figure 2B), as well as synthetic neonatal MRIs derived from a Korean neonate label image from the CU-neonates (Figure 2C). Additional synthetic cases are presented in Supplementary Information Figure S1-3. Our Fetal&Neonatal-DDPM generated high-quality synthetic MRIs for both fetal and neonatal cases. We noticed that the synthetic MRIs often surpassed the quality of original MRIs, showing features such as the choroid plexus (see yellow arrows in Figure 2), blood vessels (see purple arrows in Figure 2), and variety of surrounding tissues within the background outside of the fetal and neonatal brain. This background is either free of surrounding tissues, i.e. black (see also 'Combined pathology' in Figure 2A), or includes maternal tissues (see also 'Synthetic' in Figure 2A) or neonatal head, as prevalent in the training data. Moreover, alignment of features between the synthetic MRIs and their label images was improved compared to the original MRI-annotation pairs (see also red arrows in Figure 2), for example, the borders of the cerebral cortex always very well matched the corresponding label in the synthetic data, which is mainly due to the fact that the labels, such as the FeTA, are manual annotations, which contain anatomical inaccuracies. We noted artifacts corresponding to consistent inter-slice image intensity shifts visible in the sagittal and coronal planes of the synthetic images (see also green arrows in Figure 2).

Table 5 presents the evaluation results from four board certified radiologists rating real and synthetic data. The synthetic data, generated by our Fetal&Neonatal-DDPM model, was rated as having significantly (unpaired Student's t-test with unequal variances assumed, $p < 0.05$) greater diagnostic value than the real data, achieving an average score of 1.73 compared to 1.34 for the real data. Notably, more than 10 real images were classified as unusable, while only 1.25 synthetic images, on average, fell into this category. Rater 1, 2, and 4 consistently rated the synthetic cases as significantly higher in quality than the real cases. Overall, Rater 3 rated cases of lower quality than the other raters. The raters noted that occasionally inter-slice contrast artifacts were present, but that these did not compromise overall image quality or diagnostic value.

### 3.2. Segmentation performance

Figure 3 shows the segmentation performance on both FeTA2021$_{test}$ (Figure 3A) and specifically for Zurich-spinabifida$_{severe}$ (Figure 3B). Below we assess the performance based on different categories.

***Synthetic versus real data*** Training segmentation networks on synthetic pathological MRI-label pairs derived from healthy MRIs improved Dice score performance across both datasets, compared to training on original healthy MRIs alone. Performance increase was most notable for Zurich-spinabifida$_{severe}$ (Figure 3B), where training on S33+S198-pathologies (light orange plot in Figure 3) achieved the best Dice score performance (S33+S198-pathologies: 0.8628 vs. R33-healthy: 0.8167), especially in segmenting the cerebral ventricles (S33+S198-pathologies: 0.9253 vs. R33-healthy: 0.7317).

Training on S80-FeTA also outperformed R80-FeTA in both Zurich-spinabifida$_{severe}$ (S80-FeTA: 0.8944 vs. R80-FeTA: 0.8551) and FeTA2021test (S80-FeTA: 0.8158 vs. R80-FeTA: 0.8038). Marginal improvements were observed for S33+S198-pathologies over S80 in FeTA2021$_{test}$ (S33+S198-pathologies: 0.8176 vs. S80-FeTA: 0.8158), although outliers were more pronounced for S33+S198-pathologies and persisted for each segmentation network.

***Small and large sample sizes*** Increasing the sample sizes of the synthetic pathological training data for the FeTA dataset contributed to better overall segmentation performance when assessed in the on Zurich-spinabifida$_{severe}$ (S33+S198-pathologies: 0.8628 vs. S33+S47-pathologies: 0.8463). Conversely, for the dHCP fetal dataset, adding more synthetic cases did not improve performance, with results remaining relatively unchanged. A similar trend was observed when segmentation networks were assessed on FeTA2021$_{test}$, where S33+S198-ventriculomegaly (0.8153) outperformed S33+S47-pathologies (0.8063).

***Use of different datasets for synthetic data generation*** Using the 33 healthy cases from the FeTA2021$_{train}$ dataset to generate synthetic pathological data resulted in better overall segmentation performance compared to using 256 healthy cases from the dHCP fetal dataset, both when assessed on FeTA2021$_{test}$ (S33+S198-pathologies: 0.8176 vs. S265+S1060-pathologies: 0.7838) and on Zurich-spinabifida$_{severe}$ (S33+S198-pathologies: 0.8628 vs. S265+S1060-pathologies: 0.8323).

***Synthesizing different pathologies*** Training segmentation networks on synthetic pathological data derived from healthy data with label morphological modifications for diverse pathologies, including ventriculomegaly, cerebellum and pontocerebellar hypoplasia, and microcephaly, yielded overall marginally better performance than when including only synthetic ventriculomegaly cases. This was particularly evident when training with synthetic pathological data derived from the 33 healthy cases from the FeTA2021$_{train}$ dataset and was more pronounced when assessed on Zurich-spinabifida$_{severe}$ (S33+S198-pathologies: 0.8628 vs S33+S198-ventriculomegaly: 0.8573).

***Generalizability across structures*** When training segmentation networks on synthetic pathological ventriculomegaly data, notable improvements were observed in segmenting the cerebral ventricles, particularly in severe pathological cases of ventriculomegaly (Zurich-spinabifida$_{severe}$). Synthetic data with variations in the brainstem, i.e. pontocerebellar hypoplasia, improved the brainstem segmentation, demonstrated particularly in FeTA2021$_{test}$; however, improvements in cerebellum segmentation were not observed. We also observed a trend of improved segmentation in deep gray matter, gray matter, and white matter with the inclusion of synthetic pathological training data.

Figure 4 qualitatively shows the findings drawn from Figure 3 for a representative Zurich-spinabifida$_{severe}$ test case for all trained networks, with notable improvements in ventricle segmentations. Particularly training on S33+S198-pathologies (light orange square in Figure 4) demonstrated enhanced performance across all structures.

## 4. Discussion

This study demonstrates the potential of using synthetic pathological MRI data generated by a novel diffusion model framework, Fetal&Neonatal-DDPM, to address key challenges in fetal and neonatal MRI analysis. By generating synthetic pathological datasets through label modifications, we can overcome several limitations, such as the scarcity of annotated pathological data for training DL models. Radiologist evaluations confirmed the high quality of the synthetic data generated in this study, surpassing the quality of real MRIs. Additionally, the results highlight the ability of synthetic data to enhance segmentation accuracy, especially for pathological cases. Our framework is particularly useful for research projects where the training data or image segmentation development primarily involves healthy subjects, while the clinical application targets pathological brains for which expert ground truth annotations are unavailable or challenging to acquire.

### 4.1. Synthetic data quality and clinical utility

The synthetic MRIs generated by the Fetal&Neonatal-DDPM exhibited 'good' diagnostic quality and of significantly higher quality than real data, as evidenced by radiologist evaluations. Three radiologists assessed the synthetic data significantly as superior (Rater 1, 2, and 4), while Rater 3 generally rated all cases poorer than the other raters and assessment did not indicate an overall improved diagnostic quality than the real data. Ultimately, qualitative radiological assessments are inherently subjective. Compared to a previous pre-print from our group using GANs (Fernandez Garcia et al., 2022), we noted that the DDPM was able to synthesize more realistic and finer anatomical features, such as vascular structures, the germinal matrix, choroid plexus, and captured anatomical features, such as heterogeneous white matter patterns within labels corresponding to the same overall structure. It was also able to capture flow-voids in the dilated ventricles and other MRI related artifacts. These features, which were not represented in the input label maps, were clearly added by the generative network based on population probabilistic information, and thereby reduce the re-identification risks by such features.

The realistic pathological MRIs from modified healthy labels not only enhances the diversity and quality of the MRI datasets but also improves the visual quality of originally low-quality images, which was the case in many very young fetuses with pathological conditions, due to excessive motion. In the FeTA dataset, many pathological cases were younger than 26 weeks GA and had very blurry image quality, often marked as 'unusable' by the radiological rating. All these images were synthesized, and new image features were practically introduced at very high quality by the Fetal&Neonatal-DDPM. This likely contributed to the increased segmentation performance of the cerebellum label in the spina bifida subset of our test data. Additionally, these synthetic images encompassed enhanced segmentation accuracy between the MRI-label pairs, correcting inaccuracies in original expert annotations (see also red arrows in Figure 2). Therefore, our proposed method may be used as an image processing tool to improve quality fetal and neonatal MRI data.

Qualitative assessments across multiple datasets, including FeTA2021, dHCP fetal, and CU-neonates, demonstrate the generalizability of our proposed method. These findings highlight the model's ability to generate diagnostically valuable data and its adaptability to diverse demographic populations. This consistent performance across varied datasets suggests that this method could address gaps in MRI data quality for underrepresented groups.

### 4.2. Segmentation performance

Integrating synthetic pathological data into training state-of-the-art segmentation nnU-Net (Isensee et al., 2021) models substantially improved performance, particularly for pathological cases. These findings support the hypothesis that synthetic pathological data can compensate for the lack of pathological representation in training datasets. A key strength of this study lies in the diversity of simulated pathologies, achieved through label modifications for conditions such as ventriculomegaly, cerebellar hypoplasia, and microcephaly. The pathology generator's ability to create an infinite range of clinically relevant scenarios, including co-occurring conditions, further demonstrates its versatility and potential for enhancing model training.

An interesting finding is the improved segmentation performance of S80-FeTA compared to R80-FeTA. This may be attributed to the better alignment of the synthetic MRI visual features (anatomical borders) with those in the annotated label images (see red arrows in Figure 2) and the improved overall quality of the synthetic fetal MRIs compared to real MRIs. The higher-quality MRI-label pairs likely enabled the model to better capture the complex features and characteristics of the fetal brain, leading to more precise segmentations. This underscores the potential of synthetic data as a transformative resource for analytical tasks such as segmentation.

**Limitations and future work**

Despite the promising results, this study has limitations. First, the training time for the Fetal&Neonatal-DDPM model was relatively long (24 days). While extended training time improved network stability and helped reduce poor-quality output, it remains a time-intensive process. Second, the quality of the synthetic data was influenced by the seed of the input noise image during inference. Different noise instances generated varied training data for the same label image (see also Figure 2 and Figure S3), but occasionally, some outputs lacked intricate image features and were of lower quality compared to others generated from the same label image but with a different noise seed. This issue was more prominent in pathological data, likely due to the underrepresentation of pathological data in the training set. To address this, multiple noise instances can be generated, and the best output can be selected. Additionally, we extended the training duration beyond the initial recommendation to mitigate this issue. Future research could investigate whether extending training duration further reduces the occurrence of poor-quality outputs.

Further, we used a simplistic approach to create pathological label maps from healthy labels, which is not an accurate and comprehensive representation of the brain's anatomical changes in these conditions. For example, we dilated the ventricles, but the consequent distortion of the cortical surface (flattening of sulci) by physical forces was not modeled or simulated. We also focused on a handful of selected pathologies, which is far from the diversity that could be expected in a larger clinical center.

Third, while there was an overall performance increase when including pathological synthetic data, for some pathological cases, performance gains was not notable, as demonstrated by the outliers in Zurich-spinabifida$_{severe}$ (Figure 3B). This may be due to the heterogeneous nature of pathologies, which affect the entire brain rather than isolated structures. For instance, in severe ventriculomegaly, the extreme dilation of the ventricles may lead to malformations of all brain structures. Simulating this complex interplay using morphological modifications remains challenging. Nonetheless, our method demonstrates its ability to synthesize realistic fetal and neonatal MRIs when appropriate pathological label images are available, as demonstrated by the improved performance of S80-FeTA compared to R80-FeTA. Future research will explore alternative generative approaches to generate more diverse heterogeneous morphological label images, aiming to further improve analytical tasks such as segmentation.

## 5. Conclusion

This study highlights the potential of generative AI as a powerful data augmentation tool for enhancing the segmentation of pathological fetal MRI images, particularly in cases with complex pathologies like spina bifida or severe ventriculomegaly. By generating synthetic pathological MRIs to overcome the scarcity of annotated pathological datasets, our approach allows for the training of robust segmentation models without relying on extensive pathological data. Furthermore, this method shows promise for data anonymization, advancing clinical research and analysis, and diagnostic practices. Synthetic data generation could play an important role in overcoming data scarcity in pediatric neuroimaging and beyond.

**Acknowledgements**

This project was supported by the Swiss National Science Foundation, grant Nr. IZKSZ3_218590. Data were provided by the developing Human Connectome Project, KCL-Imperial-Oxford

Consortium funded by the European Research Council under the European Union Seventh Framework Programme (FP/2007-2013) / ERC Grant Agreement no. [319456]. We are grateful to the families who generously supported this trial. NDA study DOI: [10.15154/xth4-af39](10.15154/xth4-af39)

**Data availability statement**

Upon peer-reviewed publication, we will publicly release our Fetal&Neonatal-DDPM code, label modification framework, and an example of a fetal label image for MRI synthesis.

**Table 1:** Summary of the imaging and preprocessing characteristics of each dataset.

| | FeTA2021 | Zurich-spinabifida | Zurich-controls | CU-neonates | dHCP fetal | dHCP neonates |
|---|---|---|---|---|---|---|
| **Imaging modalities** | T2w | T2w | T2w | T1w, T2w (3D SPACE), DTI, SWI | T1w, T2w, fMRI, diffusion MRI | T1w, T2w, fMRI, diffusion MRI |
| **Scanner type** | GE 1.5T/3T (Signa Discovery MR450 and MR750) | GE 1.5T/3T (Signa Discovery MR450 and MR750) | GE 1.5T/3T (Signa Discovery MR450 and MR750) | Siemens 3T (Magnetom Vida) | Philips 3T (Achieva) | Philips 3T (Achieva) |
| **Pre-processing** | SR reconstructions (mialSR, Simple IRTK), maternal tissue excluded | SR reconstructions, semi-automated masking via MeVisLab, maternal tissue excluded | SR reconstructions, semi-automated masking via NesVOR, maternal tissue excluded | -- | Motion correction, shimming (B0, RF), SVR for 3D isotropic volumes | Motion correction, shimming (B0, RF), SVR for 3D isotropic volumes |
| **Labels** | 7 | 7 | 7 | -- | 9, 87 | 9, 87 |
| **Pathology focus** | Includes healthy cases, ventriculomegaly, spina bifida, and posterior fossa malformation. | Exclusively spina bifida cases | | Preterm neonates | -- | Includes both preterm and term-born neonates |

**Table 2:** Number of cases (*neurotypical, pathological*) and the specific age ranges for each dataset utilized in this study (*left*) and utilized for Fetal&Neonatal-DDPM training (*right*).

| Datasets | All | | | Fetal&Neonatal-DDPM training | | |
|---|---|---|---|---|---|---|
| | Neuro-typical | Pathological | Age range (weeks) | Neuro-typical | Pathological | Age range (weeks) |
| **FeTA2021** | 48 | 72 | 20-35 GA | 14 | 11 | 23-35 GA |
| **Zurich-controls** | 44 | | 20-34 GA | 20 | | 24-35 GA |
| **Zurich-spinabifida** | | 90 | 25-30 GA | | 20 | 25-30 GA |
| **dHCP fetal** | 297 | | 21-38 GA | 104 | | 21-38 GA |
| **dHCP neonates** | 558 | | 23-44 GCA | 558 | | 23-44 GCA |
| **CU-neonates** | 94 | | 29-42 GCA | | | |
| **Total** | **1041** | **162** | **20-44 GCA** | **696** | **31** | **21-44 GCA** |

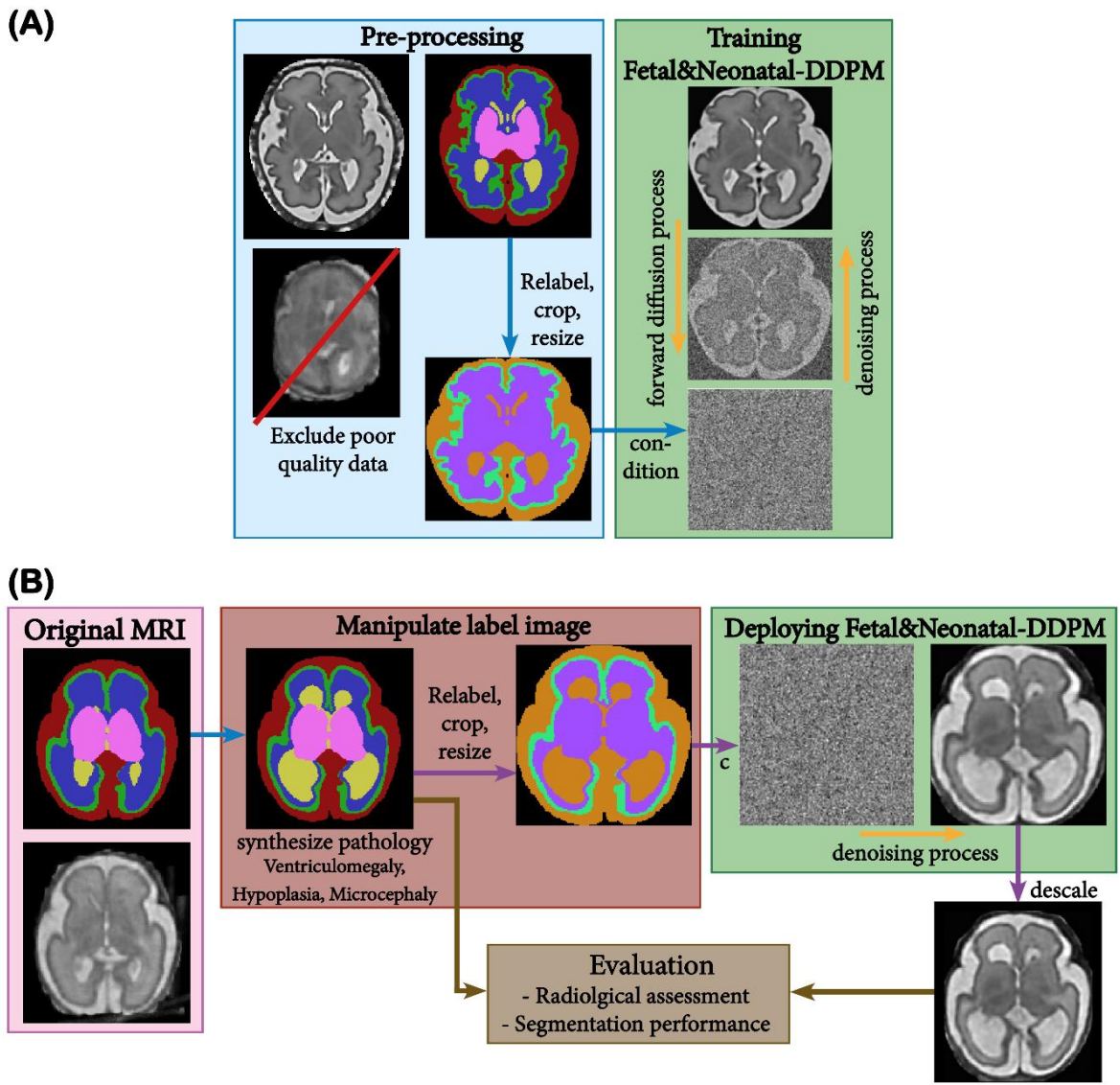

**Figure 1:** Illustrative flowchart of the primary methodological steps. **(A)** Diffusion model training. We trained our Fetal&Neonatal-DDPM diffusion model on high-quality fetal and neonatal data, excluding poor-quality data. **(B)** Pathological label and MRI synthesis. We simulated pathological label images by altering healthy ones through label morphological modifications. Ventriculomegaly is shown.

**Table 3:** Board certified details for each rater.

| Rater | Board certification | Experience in reporting MRI (years) | Experience in working with fetal or neonatal MRI (years) |
|---|---|---|---|
| 1 | Radiology | 5 | 0 |
| 2 | Radiology | 8 | 3 |
| 3 | Pediatric radiology | 15 | 11 |
| 4 | Radiology/ Neuroradiology/ Pediatric radiology | 12 | 8 |

**Table 4**: All 12 configurations of training data for nnU-Net model training. Real (R) and synthetic (S) data are used derived from the FeTA2021 Challenge training set and the dHCP fetal dataset.

|         | **Abbreviation**              | **Summary**                                                                                                                                                                                 |
|---------|-------------------------------|---------------------------------------------------------------------------------------------------------------------------------------------------------------------------------------------|
| **Exp. 1**  | R33-healthy                   | 33 healthy MRI-label pairs from the FeTA2021 Challenge training set.                                                                                                                        |
| **Exp. 2**  | S33+S47-ventriculomegaly      | 33 synthetic equivalents of R33-healthy (S33) plus 47 synthetic ventriculomegaly MRI-label pairs generated from manipulated R33-healthy label images (S47-ventriculomegaly).                |
| **Exp. 3**  | S33+S198-ventriculomegaly     | S33 plus 198 synthetic ventriculomegaly MRI-label pairs. Here, six synthetic alternatives per subject are generated.                                                                         |
| **Exp. 4**  | S33+S47-pathologies           | S33 plus 47 synthetic ventriculomegaly, hypoplasia and microcephaly MRI-label pairs generated from manipulated R33-healthy label images (S47-pathologies).                                   |
| **Exp. 5**  | S33+S198-pathologies          | S33 plus 198 synthetic ventriculomegaly, hypoplasia and microcephaly MRI-label pairs.                                                                                                        |
| **Exp. 6**  | R265-dHCP                     | 265 MRI-label pairs from the open-source dHCP fetal dataset.                                                                                                                                 |
| **Exp. 7**  | S265+S265-ventriculomegaly    | 265 synthetic equivalents of R265-dHCP (S265) plus 265 synthetic ventriculomegaly MRI-label pairs generated from manipulated R265-dHCP label images.                                         |
| **Exp. 8**  | S265+S1060-ventriculomegaly   | S265-dHCP plus 1060 synthetic ventriculomegaly MRI-label pairs generated from manipulated R265-dHCP label images. Here, four synthetic alternatives per subject are generated.               |
| **Exp. 9**  | S265+S265-pathologies         | S265 plus 265 synthetic ventriculomegaly, hypoplasia and microcephaly MRI-label pairs generated from manipulated R265-dHCP label images.                                                     |
| **Exp. 10** | S265+S1060-pathologies        | S265-dHCP plus 1060 synthetic ventriculomegaly, hypoplasia and microcephaly MRI-label pairs generated from manipulated R265-dHCP label images.                                               |
| **Exp. 11** | R80-FeTA                      | 80 healthy and pathological MRI-label pairs from the FeTA2021 Challenge training set.                                                                                                        |
| **Exp. 12** | S80-FeTA                      | 80 synthetic equivalents of R80.                                                                                                                                                             |

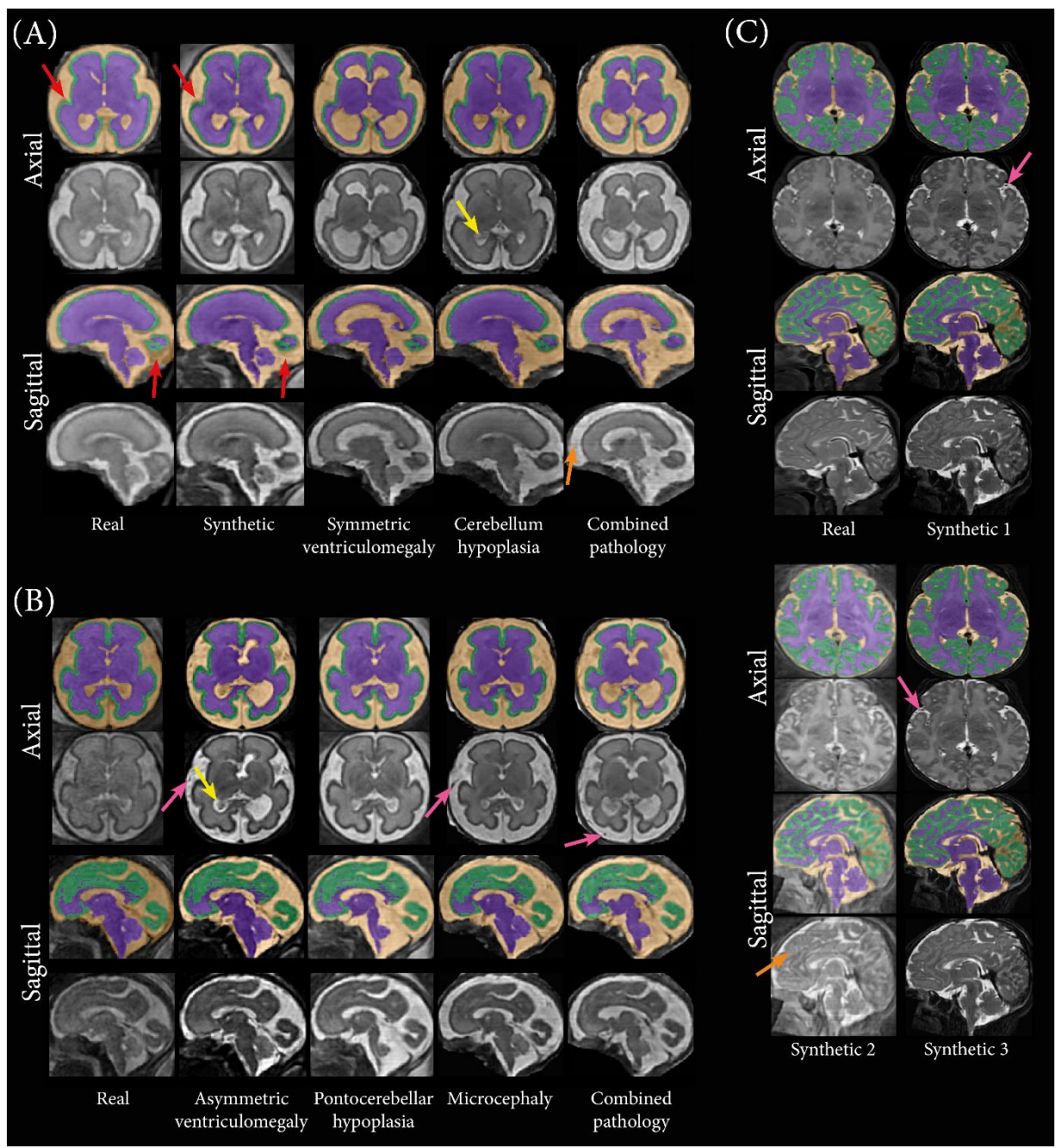

**Figure 2:** Example synthetic MRIs generated from original and modified pathological label images derived from a healthy FeTA2021$_{train}$ case of 26 weeks GA **(A)** and dHCP fetal case of 28 weeks GA **(B)** using our Fetal&Neonatal-DDPM. **(C)** presents synthetic MRIs generated from original label images from derived from a Korean neonate from the CU-neonates dataset of 38 weeks GCA. Representative axial and sagittal slices are shown. The leftmost column (or left topmost column for (C)) displays the original MRI, while subsequent columns present synthetic MRIs. Red arrows highlight areas with poor alignment in original MRI-annotation pairs but improved in synthetic MRI-label pairs. Other arrows highlight areas with synthetically generated choroid plexus (*yellow*), blood vessels in external CSF (*purple*), and line artifacts different contrast than neighboring slices (*orange*). Images are displayed in diffusion network inference dimensions (160×160×160).

**Table 5:** Summary of the quality scores for each rater.

| Rater | Unusable quality | | Poor quality | | Good quality | | Excellent quality | |
|---|---|---|---|---|---|---|---|---|
| | Real | Synthetic | Real | Synthetic | Real | Synthetic | Real | Synthetic |
| **1** | 9 | 0 | 14 | 5 | 13 | 27 | 14 | 18 |
| **2** | 15 | 1 | 10 | 8 | 14 | 32 | 11 | 9 |
| **3** | 13 | 3 | 24 | 43 | 13 | 3 | 0 | 1 |
| **4** | 6 | 1 | 15 | 8 | 20 | 32 | 9 | 9 |
| **Average** | **10.75** | **1.25** | **15.75** | **16** | **15** | **23.5** | **8.5** | **9.25** |

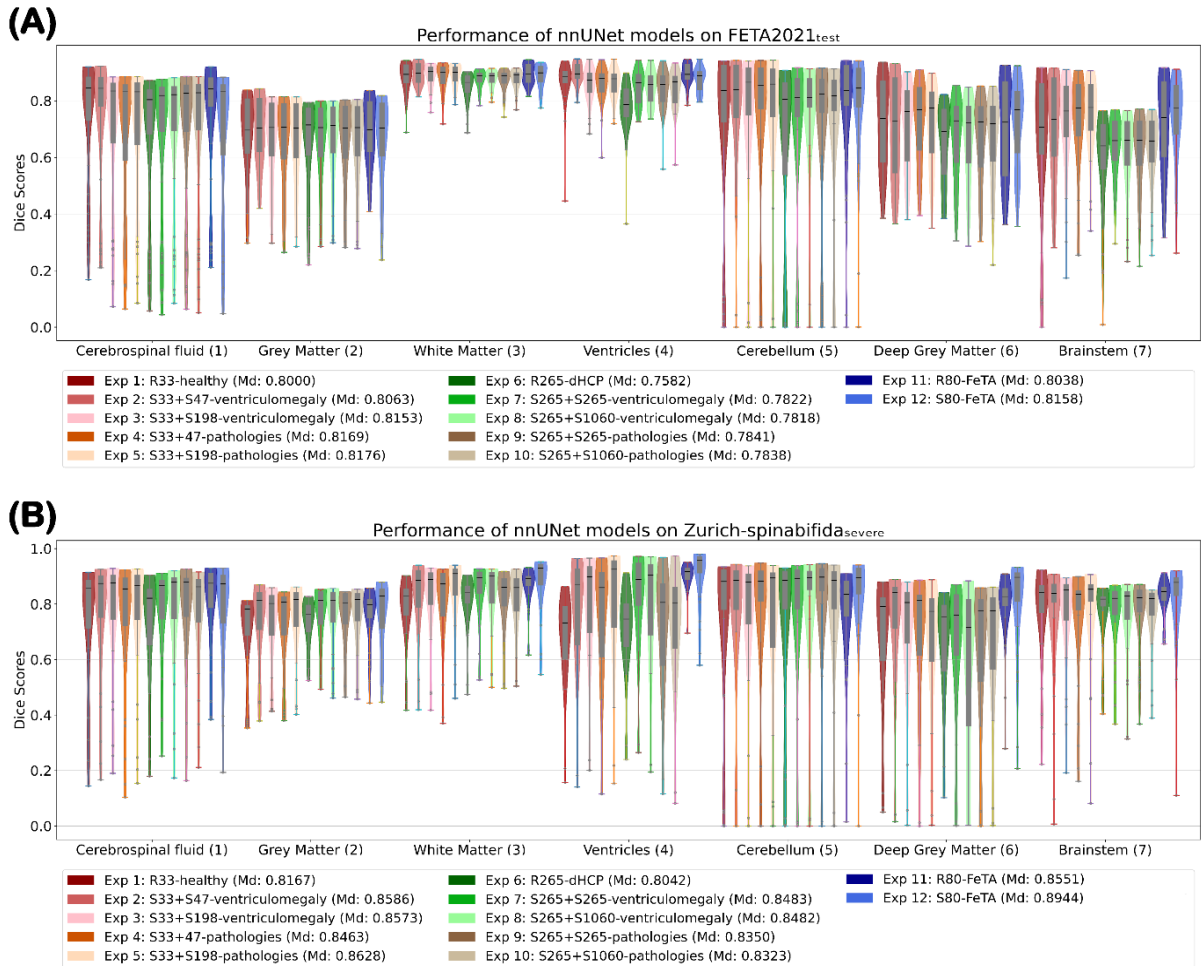

**Figure 3:** Violin plots with box-and-whisker plots overlaid of the Dice score performance for twelve nnU-Net models, trained with different configurations as described in Table 4, evaluated on the FeTA Challenge test set **(A)** and the most severe ventriculomegaly cases with spina bifida **(B)** per label. The legend shows the average of the median scores across all labels for each nnU-Net model (*in brackets*), representing overall model performance.

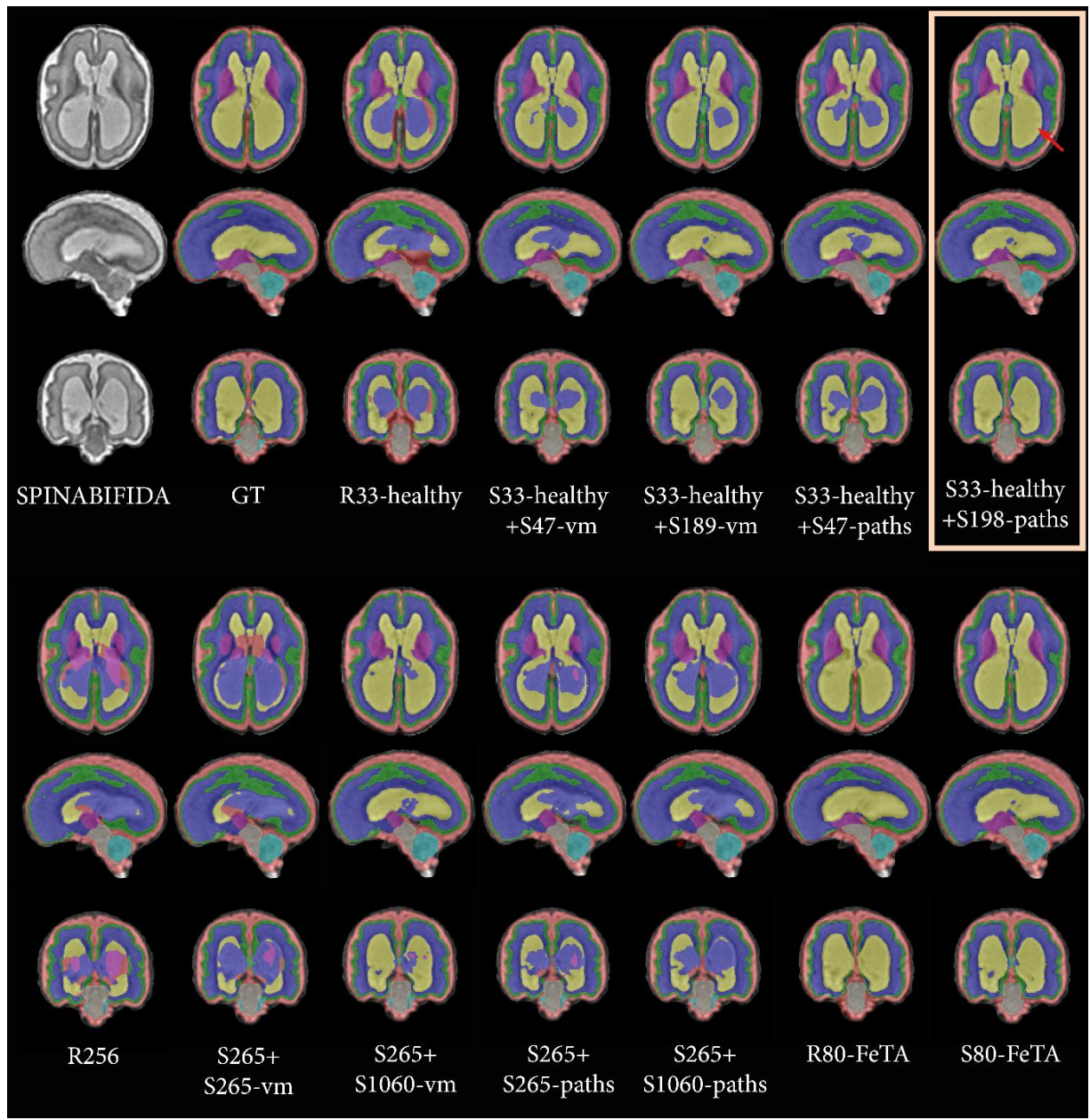

**Figure 4:** Example segmentations for a representative case (28 GA) from Zurich-spinabifida$_{severe}$, on which twelve nnUNet models, trained with different configurations as described in Table 4, were evaluated. In this figure, "ventriculomegaly" is abbreviated as "vm" and "pathologies" as "paths." Original MRIs alongside ground truth annotations are displayed at the top left. Particularly ventricular regions were correctly segmented by S33+S198-pathologies (*right top, light orange box*), highlighted by the red arrows.

*Page 1-4 is the sheet presented to the expert radiologists including examples of images.*

## MRI quality and clinical diagnostic value assessment

This assessment aims to evaluate the quality of a novel image processing approach. The form begins with a table summarizing diagnostic scores, followed by a detailed evaluation of individual cases.

The table provides four quality scores that measure the diagnostic utility of the images, rated on a scale from 0 to 3, where 1 indicates poor diagnostic utility, and 3 indicates excellent diagnostic quality. Additionally, a score of 0 is assigned to images that are entirely unusable for diagnostic purposes. To support the scoring process, guidelines and example observations are provided based on the visibility and clarity of key anatomical structures for each score. Users are encouraged to apply their clinical judgment and interpretations when assessing the images. Only one score is allowed per subject.

The detailed evaluation of individual cases comprises 100 reconstructed T2-weighted super-resolution (SR) reconstructed fetal images. Note that these appear intrinsically blurrier than original T2-weighted Single-Shot Fast Spin Echo (SSFSE) images. For each case, one representative slice from each plane (axial, coronal, and sagittal) is displayed. The images are scaled isotropically to 160x160x160, which are the dimensions of the processed images. Note that the displayed slices are randomly selected (excluding edge slices) and may not capture all key anatomical structures, such as the brainstem or deep gray matter. Furthermore, these images originate from diverse datasets, including both skull-stripped and background-inclusive data. Therefore, the background should be ignored during evaluation.

Please fill in the bottom details, and please write a general comment about the processed images after the evaluation on the final page.

Name:

Board certification: ( ) Radiology   ( ) Neuroradiology   ( ) Pediatric radiology   ( ) None

Years of experience in reporting MRI: \_\_\_\_

Years of experience with working with fetal or neonatal MRI: \_\_\_\_\_

**Table 1:** This table provides a framework for assessing the diagnostic quality of medical images, categorizing them into four levels based on clarity, resolution, and the presence of artifacts, with corresponding descriptions and examples of structural visibility.

| Score | Diagnostic Quality | Description | Example observations |
|---|---|---|---|
| 0 | Unusable quality | The image is **unusable**, with significant artifacts, noise, or distortions rendering it unsuitable for diagnostic or interpretative purposes. | Structures are poorly defined, with gray matter blending into white matter. CSF is obscured with other structures and the deep gray matter is not visible. |
| 1 | Poor quality | The image is **marginally diagnostic**, but suffers from noise and artifacts that obscure structures. | Boundaries between gray matter and white matter are unclear, CSF is visible but can overlap with other structures, and deep gray matter can be visible. |
| 2 | Good quality | The image is **useful for most diagnostic purposes**, with sufficient clarity and resolution to interpret and delineate key structures such as ventricles, gray matter and white matter, cerebellum and brainstem though lacking the detail necessary for complex or nuanced evaluations. There might be artifacts present but does not limit the delineation of identification of structures. | All key structures are visible. Gray matter can be defined from white matter. has distinct ventricles, and choroid plexus can be easily identified. The deep gray matter is recognizable and interpretable, though the sharpness of fine detail might be reduced. Deep gray matter is visible, though fine details may be less sharp. |
| 3 | Excellent Quality | The image is **fully diagnostic and optimal**, with exceptional clarity, resolution, and contrast, enabling detailed analysis of both subtle and complex features. No or very subtle artifacts are present, but not interfere in assessing the diagnostic value. | All key structures are exceptionally clear and visible. Gray matter is sharp and well-defined, with excellent contrast against white matter. CSF is perfectly outlined, with clear ventricle borders and choroid plexus, if present. The cerebellum and brainstem are detailed, with folia clearly visible, and deep gray matter shows intricate details. |

1. 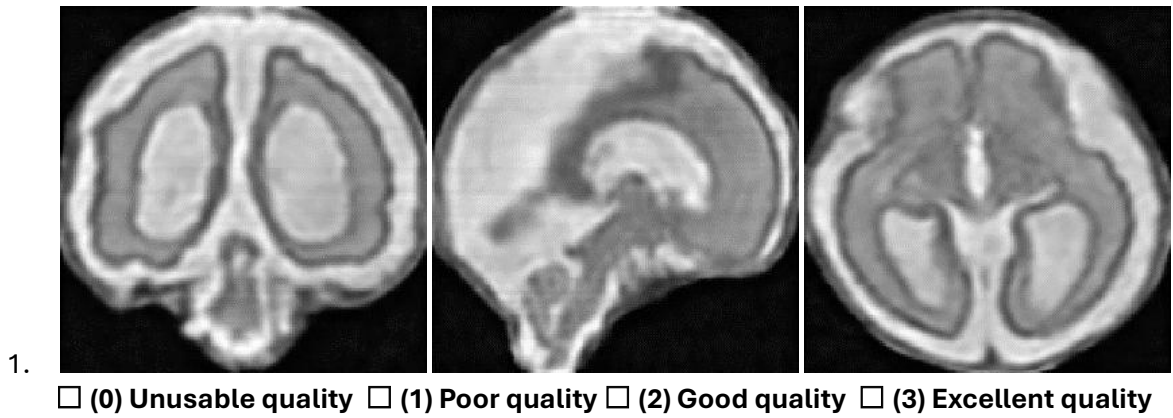
☐ **(0) Unusable quality** ☐ **(1) Poor quality** ☐ **(2) Good quality** ☐ **(3) Excellent quality**

2. 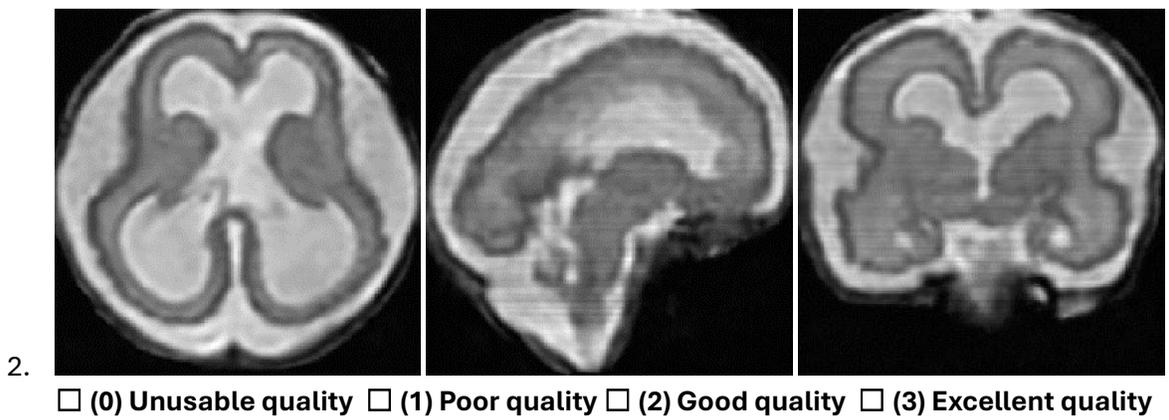
☐ **(0) Unusable quality** ☐ **(1) Poor quality** ☐ **(2) Good quality** ☐ **(3) Excellent quality**

3. 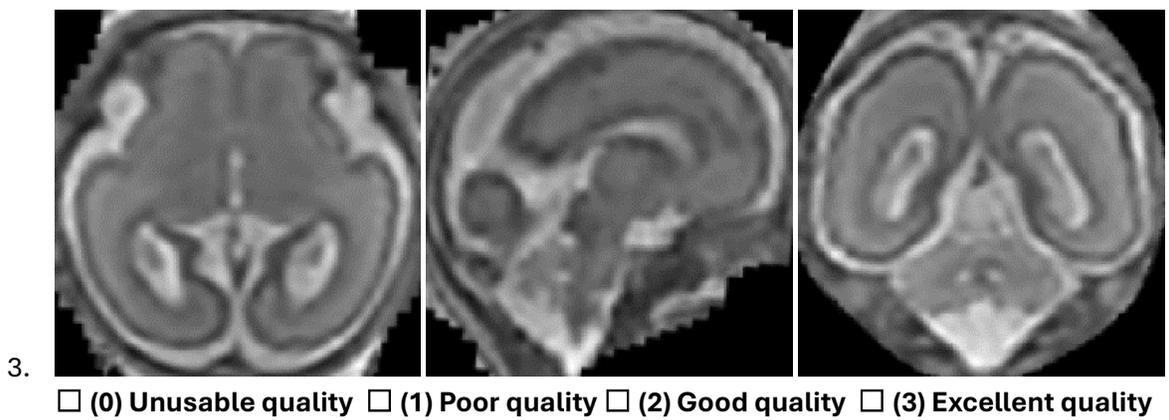
☐ **(0) Unusable quality** ☐ **(1) Poor quality** ☐ **(2) Good quality** ☐ **(3) Excellent quality**

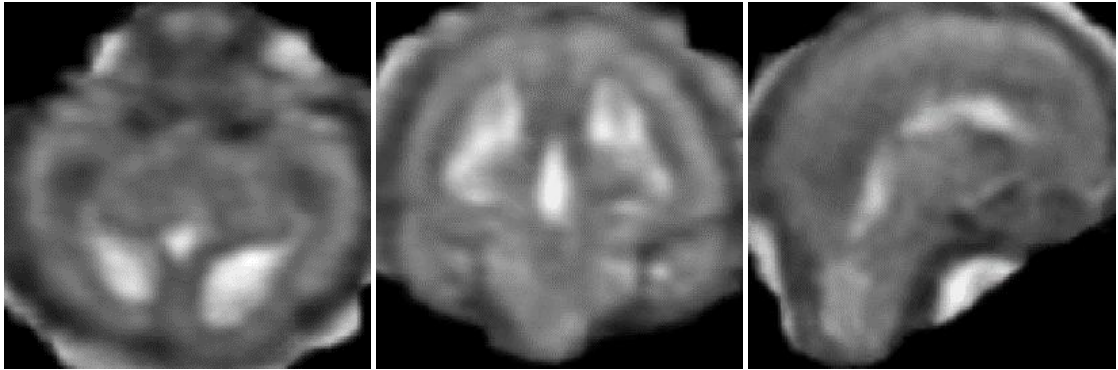

4. ☐ **(0) Unusable quality** ☐ **(1) Poor quality** ☐ **(2) Good quality** ☐ **(3) Excellent quality**

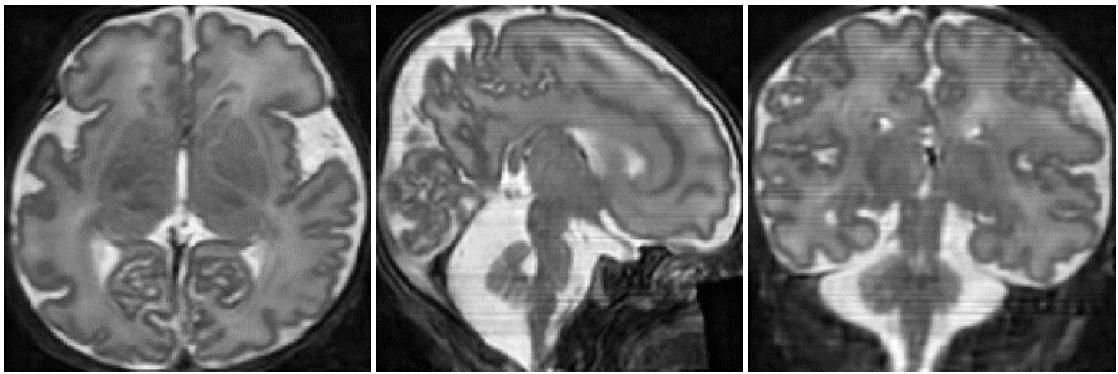

5. ☐ **(0) Unusable quality** ☐ **(1) Poor quality** ☐ **(2) Good quality** ☐ **(3) Excellent quality**

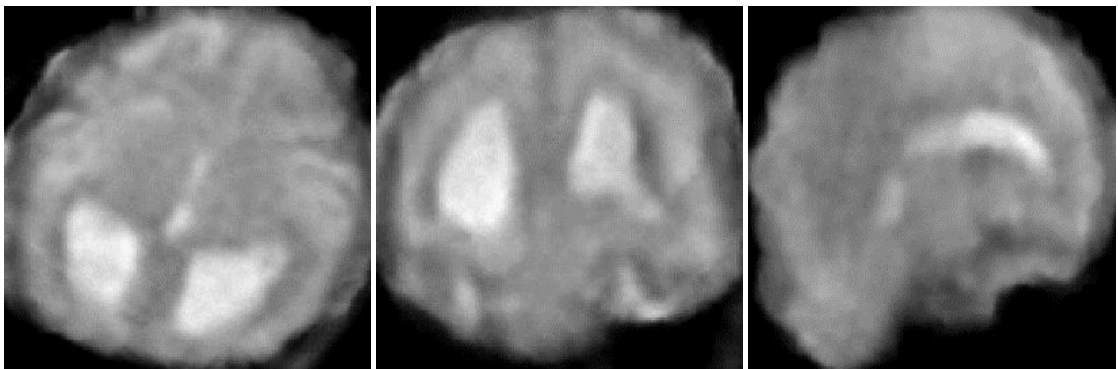

6. ☐ **(0) Unusable quality** ☐ **(1) Poor quality** ☐ **(2) Good quality** ☐ **(3) Excellent quality**

*Supporting Information for "Pathological MRI Segmentation by Synthetic Pathological Data Generation in Fetuses and Neonates" by M.P.T. Kaandorp et al. (2025).*

Below are additional examples of synthetic images for c case from the FeTA2021$_{train}$ (**Figure S1**), dHCP fetal (**Figure S2**), and CU-neonates (**Figure S3**) using our Fetal&Neonatal-DDPM.

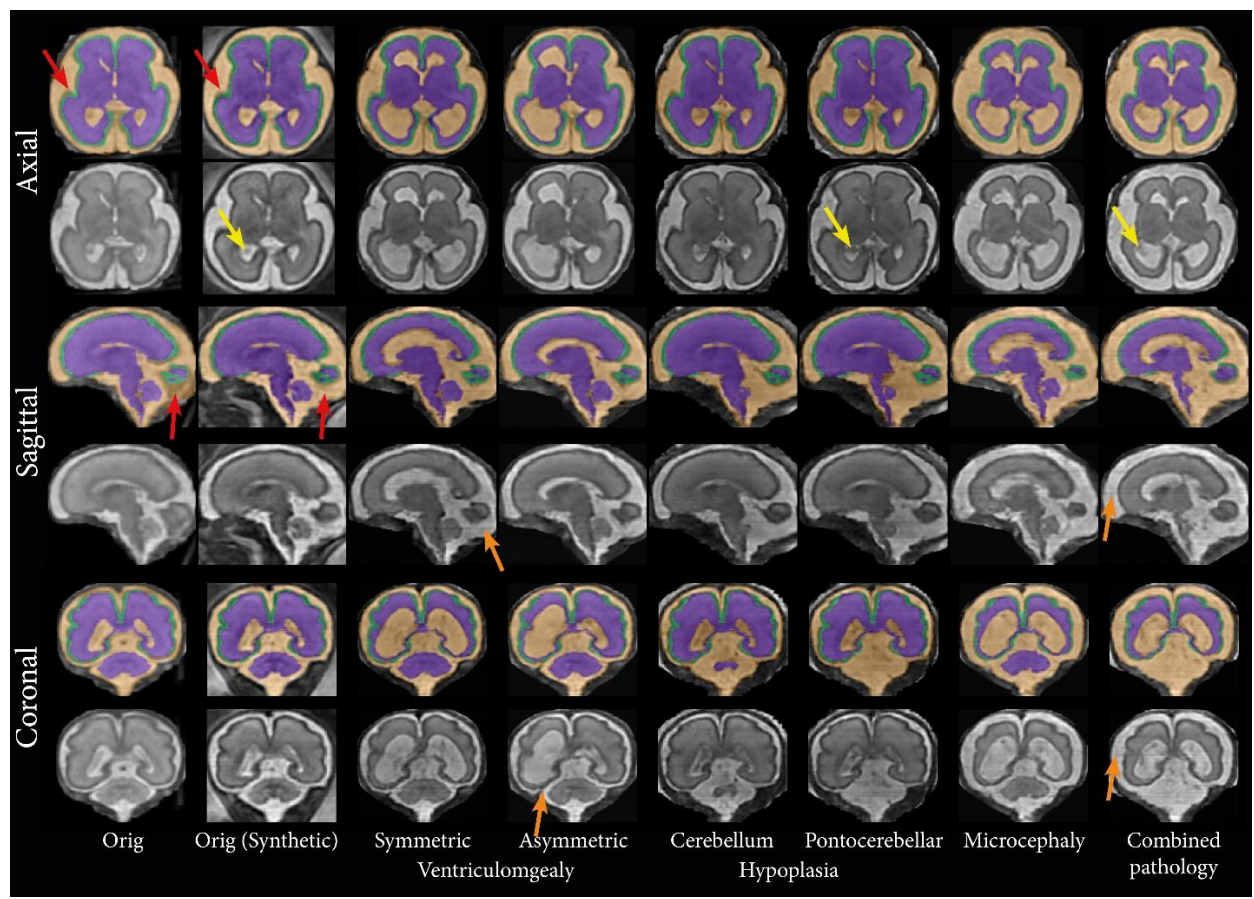

**Figure S1:** Example synthetic MRIs generated from original and modified pathological label images derived from a healthy MRI (26 weeks GA) from the FeTA challange training set using our Fetal&Neonatal-DDPM. Representative axial, coronal, and sagittal slices are shown. The leftmost column displays the original MRI, while subsequent columns present synthetic MRIs. Red arrows highlight areas with poor alignment in original MRI-annotation pairs but improved in synthetic MRI-label pairs. Other arrows highlight areas with synthetically generated choroid plexus (*yellow*) and line artifacts different contrast than neighboring slices (*orange*). Images are displayed in diffusion network inference dimensions (160x160x160).

*Supporting Information for "Pathological MRI Segmentation by Synthetic Pathological Data Generation in Fetuses and Neonates" by M.P.T. Kaandorp et al. (2025).*

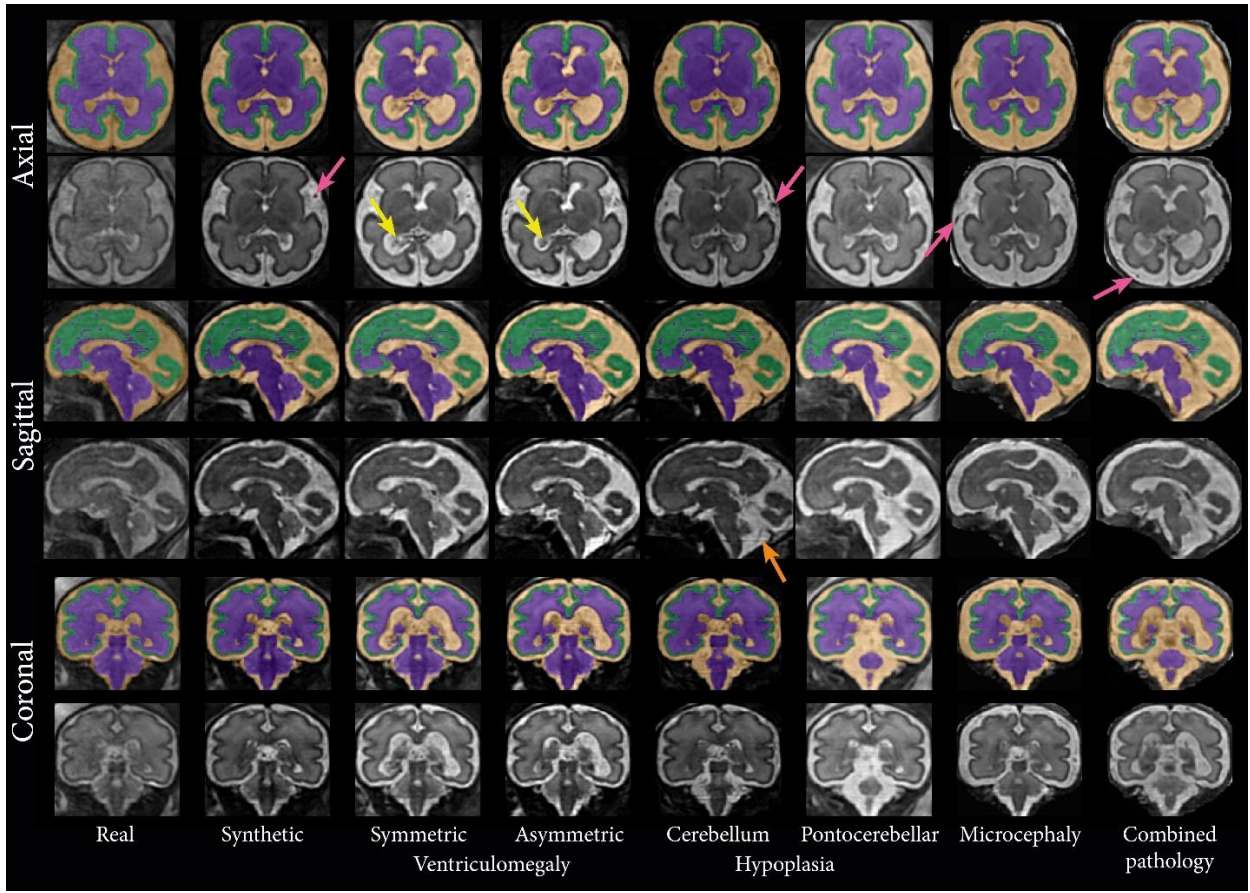

**Figure S2:** Example synthetic MRIs generated from original and modified pathological label images derived from an MRI of the dHCP fetal (28 weeks GA) dataset using our Fetal&Neonatal-DDPM. Representative axial, coronal, and sagittal slices are shown. The leftmost column displays the original MRI, while subsequent columns present synthetic MRIs. The arrows highlight areas with synthetically generated choroid plexus (*yellow*), blood vessels in external CSF (*purple*), and line artifacts different contrast than neighboring slices (*orange*). Images are displayed in diffusion network inference dimensions (160x160x160).

*Supporting Information for "Pathological MRI Segmentation by Synthetic Pathological Data Generation in Fetuses and Neonates" by M.P.T. Kaandorp et al. (2025).*

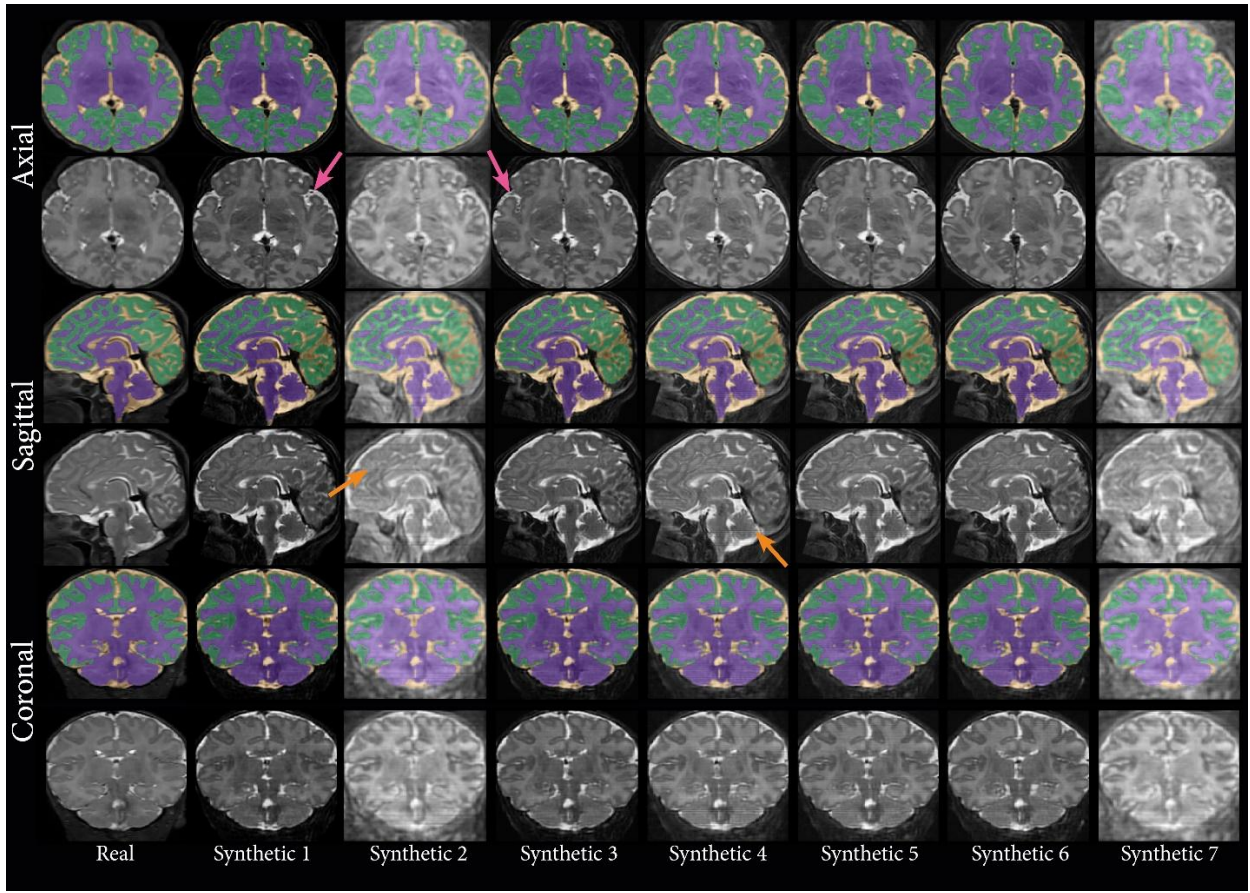

**Figure S3:** Example synthetic MRIs generated from original and modified pathological label images derived from an MRI of a Korean neonate from the CU-neonates dataset (38 weeks GCA, scanned at 1 week of age) using our Fetal&Neonatal-DDPM. Representative axial, coronal, and sagittal slices are shown. The leftmost column displays the original MRI, while subsequent columns present synthetic MRIs. The arrows highlight areas with synthetically generated blood vessels in external CSF (*purple*) and line artifacts different contrast than neighboring slices (*orange*). Images are displayed in diffusion network inference dimensions (160x160x160).